\documentclass[12pt]{article}
\usepackage{array}
\usepackage{amsmath}
\newcommand{\be}{\begin{equation}}
\newcommand{\ee}{\end{equation}}
\newcommand{\bea}{\begin{eqnarray}}
\newcommand{\eea}{\end{eqnarray}}

\begin{document}

\title{The Canonical Structure of Bigravity\footnote{This article is based on the talk given at the Fourth Zeldovich 
meeting, an international conference in honor of Ya. B. Zeldovich held in Minsk, Belarus on September 7--11, 2020. }}

\author{{V.O. Soloviev}}

\maketitle

\begin{abstract}
This work is motivated by an intention to make the theory of bigravity more comprehensible. Bigravity is a modification of the General Relativity (GR), maybe even the most natural one because it is  based on the equivalence principle. The Hamiltonian formalism in tetrad  variables transparently demonstrates the structure of bigravity 
\end{abstract}

\section{Introduction}
 Lagrangian of the bigravity is a sum of two GR Lagrangians formed of two spacetime metrics $f_{\mu\nu}$, $g_{\mu\nu}$ and a potential of their interaction
discovered by de Rham, Gabadadze, and Tolley~\cite{dRGT,Reviews}. 
The potential of bigravity simplifies when the action is expressed through tetrads~\cite{HiRo}, not metrics.  Both two sets of lapse-and-shift variables appear linearly in the Hamiltonian and can be treated as Lagrange multipliers  at primary constraints. As the theory is explicitly invariant only under diagonal diffeomorphisms of the spacetime manifold and  diagonal rotations of the spatial triads, the number of arbitrary Lagrange multipliers is 7 (1+3+3) the same is  the number of  the first class constraints.  Other Lagrange multipliers provide 10 second class constraints. The  compatibility of the primary second class constraints with  the dynamical equations provides 10 new equations, where 6 of them are the so-called tetrad symmetry conditions, and the other 4 are equivalent to the second class constraints of the metric approach. One of these 4 constraints accompanied by a corresponding primary constraint serves to remove the ghost degree of freedom.
Three other constraints together with their  (fixed by the compatibility conditions) Lagrange multipliers serve to supplement the two Hamiltonian-like constraints. This reorganization of constraints reproduces the results of the celebrated Hassan-Rosen transform~\cite{HaRo}. From the geometrical viewpoint, it is interesting to notice that three bilinear combinations of the pair of triads corresponding to the pair of spatial metrics appear in the dRGT potential. One of these combinations is symmetric and therefore can be treated
as a new (hybrid) spatial metric. Its role in the coupling to matter fields requires
a detailed investigation. It is interesting to mention the correspondence of this
combination to the spatial components of the geometric mean of the two spacetime metrics introduced by Kocic~\cite{Kocic}.

Below for spacetime coordinate indices running from 0 to 3, we
use small Greek letters; for internal indices running from 1 to 3, we
use small Latin letters from the beginning of the alphabet. For spatial indices small letters from the middle of the alphabet are used, for internal indices running from 0 to 3 the capital Latin letters are used. We deal with metrics that have common timelike and spacelike vectors. The variables related to the  metric $g_{\mu\nu}$ are marked by an upper bar.

\section{Lagrangian, potential, and variables}
From the mathematical point of view,  GR looks much simpler when expressed in the geometrical language, i.e. in variables having an evident geometrical meaning. This is  true also for  the bigravity.
The Lagrangian of bigravity is equal to the sum of two copies of the GR Lagrangian minus an interaction term called the potential
\be
{\cal L}^{(f)}=\frac{1}{16\pi G^{(f)}}\sqrt{-f}f^{\mu\nu}R_{\mu\nu}^{(f)}+{\cal L}_M^{(f)}(\psi^A, f_{\mu\nu}),\label{eq:Lf}\
\ee
\be
{\cal L}^{(g)}=\frac{1}{16\pi G^{(g)}}\sqrt{-g}g^{\mu\nu}R_{\mu\nu}^{(g)}+{\cal L}_M^{(g)}(\phi^A, g_{\mu\nu}),\label{eq:Lg}
\ee
\be
{\cal L}={\cal L}^{(f)}+{\cal L}^{(g)}
-\frac{m^2}{2\kappa}\sqrt{-g}U( f_{\mu\nu}, g_{\mu\nu}).\label{eq:L_bi}
\ee
The diffeomorphism invariance requires
\be
U( f_{\mu\nu}, g_{\mu\nu})=U(\mbox{invariants of}\  \mathsf{Y}),\qquad \mbox{where}\qquad\mathsf{Y}=g^{-1}f\equiv g^{\mu\alpha}f_{\alpha\nu}.
\ee
The first formulation of the GR in the Hamiltonian language was given by Dirac~\cite{Dirac}. If we apply Arnowitt-Deser-Misner~\cite{ADM} (ADM) variables (lapses $N$, $\bar N$, shifts $N^i$, $\bar N^i$ and induced metrics $\eta_{ij}$, $\gamma_{ij}$) and introduce a basis for spacetime tensors $(n^\alpha,e^\alpha_i)$ introduced by Kucha\u{r} and York~\cite{York} (formed by one of  metrics, let it be $f_{\mu\nu}$)  we obtain 
\be
\mathsf{Y}=g^{-1}f=u^{-2}\left(\begin{array}{cc} -[n^\mu n_\nu] & u^i[n^\mu e_{\nu i}] \\
u^j[e^\mu_j n_\nu] & \left(-u^iu^j+u^{2}\gamma^{ij}\right)[e^\mu_i e_{\nu j}] \\ \end{array}
\right), \label{eq:Ymatrix}
\ee
where 
\be
u=\frac{\bar N}{N},\qquad u^i=\frac{\bar N^i-N^i}{N}
\ee
A standard longstanding problem of the nonlinear massive gravity (and also bigravity) was the Boulware-Deser~\cite{BD} ghost arising due to nonlinearity of $\sqrt{-g}U$
in the auxiliary variable $u$.
The potential proposed by de Rham, Gabadadze, and Tolley~\cite{dRGT,Reviews} (dRGT) is as follows
\be
U=
\sum_{n=0}^4\beta_ne_n(X), \qquad \mathsf{X}=\sqrt{\mathsf{Y}},\qquad \mathsf{Y}=||g^{\mu\alpha}f_{\alpha\nu}||,
\ee
where the symmetric polynomials  of  matrix 
$
\mathsf{X}^\mu_\nu=\sqrt{||g^{-1}f||}^\mu_\nu
$ written through traces of it and its powers are the following
\begin{eqnarray}
e_0&=&1,\nonumber\\
 e_1&=&\mathrm{Tr}X,\nonumber\\
e_2&=&\frac{1}{2}\left((\mathrm{Tr}X)^2-\mathrm{Tr}X^2 \right),\nonumber\\
e_3&=&\frac{1}{6}\left((\mathrm{Tr}X)^3-3\mathrm{Tr}X\mathrm{Tr}X^2+2\mathrm{Tr}X^3 \right),\nonumber\\
e_4&=&\det X .\nonumber
\end{eqnarray}
Then a solution of the theory equations is given should give two spacetime metric tensors and all the matter fields.

\section{Kucha\u{r}'s notations}
The Hamiltonian formalism of GR becomes more transparent when given in the embedding variables, i.e. as a dynamics of hypersurfaces.  The suitable variables are the induced metric $\gamma_{ij}$  and the external curvature tensor   $K_{ij}$. In the ADM variables, the time components of the metric $g_{0\mu}$  are replaced by  the lapse and shift variables $N,N^i$ that connects the close  hypersurfaces.  In the  Kucha\u{r} approach~\cite{Kuchar} (see also York~\cite{York})    $N,N^i$ are  components of the 4-vector connecting observer positions on the closest hypersurfaces
\be
N^\alpha\equiv \frac{\partial X^\alpha}{\partial t}=Nn^\alpha+N^ie^\alpha_i,
\ee
where two coordinate frames  $X^\alpha$ and   $(\tau,x^i)$ are used. The embedding functions  $e^\alpha(\tau,x^i)$ provide one-to-one  map $
X^\alpha=e^\alpha(\tau,x^i)
$.  The three tangential to a hypersurface vectors are $e^\alpha_i=\frac{\partial e^\alpha}{\partial x^i}$.
 In bigravity we have two  unit normal vectors they are denoted as $n^\alpha$, 
    $\bar{n}^\alpha$, and they satisfy equations:
\begin{alignat}{2}
&g_{\mu\nu}\bar n^\mu\bar n^\nu=-1, \qquad &g_{\mu\nu}\bar n^\mu e^\nu_i=0, 
\nonumber\\
&f_{\mu\nu} n^\mu n^\nu=-1, \qquad &f_{\mu\nu}n^\mu e^\nu_i=0. 
\nonumber
\end{alignat}  
The canonical variables are the two induced metrics  $\eta_{ij}=f_{\mu\nu}e^\mu_ie^\nu_j$, $\gamma_{ij}=g_{\mu\nu}e^\mu_ie^\nu_j$ and the two external curvature tensors  $K_{ij}=-e^\alpha_in_{\alpha;\beta}e^\beta_j$, and  $\bar K_{ij}=-e^\alpha_i{\bar n}_{\alpha;\beta}e^\beta_j$. 
Two spacetime metrics  in their local bases $(\bar n^\alpha,e^\alpha_i)$, and $(n^\alpha,e^\alpha_i)$ are
\be
g_{\mu\nu}=-\bar n_\mu \bar n_\nu+\gamma_{ij}\bar e_\mu^i \bar e_\nu^j,\qquad f_{\mu\nu}=-n_\mu n_\nu+\eta_{ij}e_\mu^i e_\nu^j.
\ee

\section{Tetrads}
The metric tensor is not a unique choice  of a dynamical variable. There is another  possibility of the geometrical description provided by a field of orthonormal bases or tetrads~\cite{HamTetrads}. In bigravity, we  have  two such bases $F^A_\mu$, $E^A_\mu$ given at each spacetime point. The potential now can be  expressed explicitly, i.e. it is possible to find  a matrix square root of  the mixed tensor $\mathsf{Y}^\alpha_\beta=g^{\alpha\mu}f_{\mu\beta}$. But the physical content of the metric and tetrad formulations is the same only if  symmetry conditions for the tetrads are fulfilled.
The vierbeins (or tetrads) are the square root of metric
\be
g=E^TE, \qquad g_{\mu\nu}=E_{\mu A}E^A_\nu,
\ee
\be
g^{-1}=E^{-1}(E^{-1})^T,\qquad g^{\mu\nu}=E^\mu_AE^{A\nu},
\ee
Then we can extract the square root of the matrix $\mathsf{Y}$
\be
\mathsf{X}=\sqrt{g^{-1}f}=\sqrt{E^{-1}(E^{-1})^TF^TF}=E^{-1}F^T,
\ee
if  symmetry conditions are fulfilled
\be
(FE^{-1})^T=FE^{-1}.
\ee

There is a diagonal Lorentz symmetry generated by
\begin{equation}
L^+_{A B}=\left(\begin{array}{cc} 0 & L^+_{0b} \\
L^+_{a0} &L^+_{ab} \\ \end{array}
\right), \nonumber
\end{equation}
We can sacrifice $L^+_{a0}$  to achieve the null tetrad gauge for $E_{A\mu}$
\be
E_{0\mu}=\bar n_\mu.
\ee
Then the dynamical variables occur triads $e^a_i$. But we can apply this gauge only to one tetrad as the potential is invariant under diagonal tetrad rotations  
\be
F'^A_\mu=\Lambda^A_{\ B}F^B_\mu, \qquad E'^A_\mu=\Lambda^A_{\ B}E^B_\mu.
\ee
 In the article by Hinterbichler and Rosen~\cite{HiRo}, it was suggested to parametrize the additional degrees of freedom by adding an arbitrary boost transformation to the triad basis.
The  parametrization of a boost
  \begin{equation}
\Lambda^A_{ \ B}=\left(\begin{array}{cc} \varepsilon & \varepsilon v_b \\
\varepsilon v^a &{\cal P}^a_b \\ \end{array}
\right), \quad {\cal P}^a_b =\delta^a_{ \ b}+\frac{\varepsilon^2}{\varepsilon+1}v^a v_b\nonumber \ ,
\end{equation}
allows  taking the  second tetrad $F_{A\mu}$ in the form
\be
F^A_{\mu}=\Lambda^A_{ \ B}{\cal F}^B_\mu
\ee
where ${\cal F}^B_\mu$ is a second tetrad given in the time gauge.
In this work, we take parameters of this boost as canonical variables and introduce new momenta conjugate to them. This is different from the approach taken in H-R.
Therefore we get  21 pairs of canonically conjugate variables:
\be
(e_{ai},\pi^i_a),\ (\tilde f_{ai},\Pi^i_a), (\tilde v_i,\Pi^i_0),
\ee
where 
\be
\tilde f_{ai}={\cal P}_{ab}f_{bi},\qquad \tilde v_i=\tilde f_{ai}v_a.
\ee
The other variables are Lagrange multipliers $
N,N^i,u,u^i,\lambda^+_{ab},\lambda^-_{ab},\lambda^a$. The Hamiltonian is as follows
\be
\mathrm{H}=\int d^3x \Biggl[N\left( {\cal R}''+u{\cal S}'+u^i{\cal S}_i\right)+ N^i{\cal R}_i+\lambda^+_{ab}L^+_{ab}+ \lambda^-_{ab} L^-_{ab}+\lambda^aL_{a0}\Biggr].
\ee
It is necessary to compare the approach used here with the earlier work by Alexandrov~\cite{Alex} and  the preceding article~\cite{Krasnov}.  There the formalism was  developed for the two general tetrads and  two general  connections.  The spatial triads  appear  as a solution to the second class constraints. These second class constraints arise because of applying the first order Palatini formalism where connections and tetrads are initially treated as  independent variables. A common feature of  both approaches  is that the tetrad symmetry conditions appear as a consequence of the compatibility of the primary constraints with the Hamiltonian dynamics. But this is not the case for  the earlier work~\cite{Krasnov}. The tetrad approach was also considered in article~\cite{Kluson}.

\section{Implicit functions used in metric approach}
It is impossible to express the dRGT potential as an explicit function of the metric variables, therefore implicit functions are used. After extracting lapses and shifts  of  both metrics  and making a special transform  of variables~\cite{HaRo} it is possible to express the potential as a function of $3\times 3$-matrix $D^i_{\ j}$. This matrix is to be symmetrical 
\be
D^{ij}=D^{ji}, 
\ee
and satisfy the following equation
\be
\gamma^{ij}=D^i_{\ k}v^kD^j_{\ m}v^m+\varepsilon^{-2}D^{ik}D_k^{\ j}.\label{eq:D}
\ee
The above equations for $D^i_{\ j}$ follow from Eq.(\ref{eq:Ymatrix})
when the {Hassan-Rosen} transform of variables 
\be
u^i=v^i+uD^i_{\  j}v^j,\qquad \varepsilon^{-1}=\sqrt{1-\eta_{ij}v^iv^j}.
\ee
is applied.  We start from a definition of $D^i_{\ j}$ by the following formula
\be
\mathsf{X}=\sqrt{\mathsf{Y}}=\varepsilon u^{-1}\left(\begin{array}{cc} -[n^\mu n_\nu] &  v^i[n^\mu e_{\nu i}] \\
v^j [e^\mu_j n_\nu] & \left(- v^iv^j+\varepsilon^{-2}u D^{ij}\right)[e^\mu_i e_{\nu j}] \\ \end{array}
\right). \label{eq:Xmatrix}
\ee
After squaring  matrix $\mathsf{X}$ and comparing the result with the previously obtained expression for matrix $\mathsf{Y}$ we obtain  equations (\ref{eq:D}).
Therefore $D^i_{\ j}$ depends on $\eta_{ij}$, $\gamma_{ij}$ and $v^i$,
 indices of $D^i_{\ j}$ are moved up and down by $\eta_{ij}$ and its inverse $\eta^{ij}$.
After heavy calculations, it occurred possible to find expressions for derivatives of   $D^i_{\ j}$ 
with respect to canonical coordinates  $\eta_{ij}$, $\gamma_{ij}$. This allowed to calculate Poisson brackets of the potential with the other terms of the Hamiltonian and to get the constraints algebra~\cite{H-L}.

In another approach~\cite{Comelli2012,Sol} the potential as a whole is  considered as an implicit function of lapses, shifts, and induced metrics.  It is shown that if  this function fulfills the homogeneous Monge-Ampere equation in lapses and shifts then the theory is  free of the Boulware-Deser ghost. Also, it is supposed that the rank of the corresponding matrix is equal to three.  The important properties  of the implicit solutions of the Monge-Ampere equation are given in~\cite{Leznov}.

By putting to zero variables $n^i$, and so discarding the Hassan-Rosen transformation, but preserving $D^i_{\ j}$ one may arrive at the precursor theory for the Minimal Theory of Bigravity~\cite{DeFelice}.

\section{Algebra of constraints and degrees of freedom}
In this section, we present the results of the Poisson brackets calculations. With the tetrad variables, we obtain that the Hassan-Rosen transform can be written as follows
\be
u^i=v^i+u{\bar v}^i,
\ee
where
\be
v^i=f^{ia}v_a, \qquad {\bar v}^i=e^{ia}v_a.
\ee
Then we introduce
\be
{\cal S}={\cal S}'+\bar v^i{\cal S}_i,
\ee
\be
{\cal R}={\cal R}''+ v^i{\cal S}_i+u{\cal S}.
\ee
For the 1st class constraints ${\cal R},{\cal R}_i,L^+_{ab}$ we get 
\bea
\{{\cal R}(x),{\cal R}(y)\}&=&\left(
\eta^{ik}{\cal R}_k+{uu^i}{\cal S}
\right)(x)
\delta_{,i}(x,y)-\left(
x\leftrightarrow y
\right),\nonumber\\
\{{\cal R}_i(x),{\cal R}(y)\}&=&{\cal R}(x)\delta_{,i}(x,y)+u_{,i}{\cal S}\delta(x,y),\nonumber\\
\{{\cal R}_i(x),{\cal R}_j(y)\}&=&{\cal R}_j(x)\delta_{,i}(x,y)-{\cal R}_i(y)\delta_{,j}(y,x),\nonumber 
\eea
and
\bea
\{L^+_{ab}(x),L^+_{cd}(y)\}&=&\left(\delta_{ac}L^+_{db}+\delta_{bc}L^+_{ad}-\delta_{ad}L^+_{cb}-\delta_{bd}L^+_{ac}\right)\delta(x,y)\nonumber\\
\{{\cal R}_i(x),{L}^+_{ab}(y)\}&=&{L}^+_{ab}(x)\delta_{,i}(x,y)\approx 0,\nonumber\\
\{{\cal R}(x),{L}^+_{ab}(y)\}&=&0.\nonumber
\eea
For the 2nd class constraints   ${\cal S}$, $\Omega$ results are the following
\bea
\{{\cal S}(x),{\cal S}(y)\}&=&{\bar v}^i{\cal S}(x)\delta_{,i}(x,y)-{\bar v}^i{\cal S}(y)\delta_{,i}(y,x),\nonumber\\
\{{\cal R}(x),{\cal S}(y)\}&=&(u^i+u\bar{v}^i){\cal S}(x)\delta_{,i}(x,y)+\left({u}(\bar{v}^i{\cal S})_{,i}-\Omega\right)\delta(x,y),\nonumber\\
\{{\cal S}(x),\Omega(y)\}&\ne &0.\nonumber
\eea
The Hassan-Rosen transform may also be written as follows
\be
\bar{N}^i=N^i+Nv^i+\bar{N}{\bar v}^i,
\ee
The constraints $L^-_{ab}$, $G_{cd}$, $L_{a0}$ are second class as we get
\begin{align}
\{L^-_{ab}(x),G_{cd}(y)\}&=\left[\delta_{ac}z_{(bd)}-\delta_{ad}z_{(cb)}-\delta_{bc}z_{(ad)}+\delta_{bd}z_{(ca)}\right]\nonumber\\
&\times \delta(x,y)\ne 0,\nonumber\\
\{L_{a0}(x),{\cal S}_i(y)\}&=e\tilde{f}_{bi}\left[\beta_1\delta_{ba}e_0(z)+\beta_2(\delta_{ba}e_1(z)-z_{ba})\right.\nonumber\\
&\left.+\beta_3(\delta_{ba}e_2(z)+z_{bc}z_{ca}-zz_{ba})\right]\delta(x,y)\ne 0.\nonumber
\end{align}
where
\be
z_{ab}=e_{ai}\tilde f^{ib}=z_{ba},\qquad \tilde f^{ib}={\cal P}^{-1}_{ab}f^{ia},\qquad
 {\cal P}^{-1}_{ab}=\delta_{ab}-\frac{\varepsilon}{\varepsilon +1}v_av_b.
\ee


\section{Conclusion}
The results of this work are summarized in Table 1 where the number of gravitational degrees of freedom is calculated according to the  formula
\be
{\mathrm{DOF}}=\frac{1}{2}\left(n-2n_{f.c.}-n_{s.c.} \right). 
\ee
The advantages of the proposed approach are the following.
 The potential (and so the Hamiltonian) is linear  in the lapses and shifts $N$, $\bar{N}$, $N^i$, $\bar{N}^i$.
All the nondynamical functions are Lagrange multipliers.
The tetrad symmetry conditions follow from the Dirac procedure. The first three of them appear as  secondary constraints, and the other three as a fixing of the Lagrangian multiplier $u^i$.
 Therefore the Hassan-Rosen transform  is derived, and not postulated.
 Neither implicit functions, nor Dirac brackets are involved in the calculations.
  The geometrical meaning of the coefficients  standing in the algebra of constraints is uncovered.

We hope that the obtained results may be applied to
the Cauchy problem,
 the perturbation theory,
 the numerical bigravity,
 the canonical quantization of bigravity.

The author is  most grateful to the Organizing Committee of the IV Zeldovich Meeting for the opportunity to participate in this exciting  conference and to the participants for their attention and discussion. 
\begin{table}
\begin{tabular}{|c||c|c|c|}
\hline 
  & BiGrav (general) & BiGrav (dRGT) & BiGrav (vierbein) \\
\hline\hline
 $(q,p)$ &  $(\gamma_{ij},\pi^{ij}),(\eta_{ij},\Pi^{ij})$ & $(\gamma_{ij},\pi^{ij}),(\eta_{ij},\Pi^{ij})$ & $(e_{ia},\pi^{ia}),(\tilde{f}_{ia},\Pi^{ia})$\\
 & & &$(p_i,\Pi^i_0)$ \\
$n$ & 24 & 24 &   42 \\
\hline
1st class &  ${\cal R},{\cal R}_i$ & ${\cal R},{\cal R}_i$ & ${\cal R},{\cal R}_i,L^+_{ab}$ \\
$n_{f.c.}$ & 4 & 4 & 7 \\
\hline
 2nd class & ---  &${\cal S},\Omega$&  ${\cal S},\Omega,L^-_{ab},G_{ab}$ \\
 & & &$L_{a0},{\cal S}_i$ \\
$n_{s.c.}$ & 0 & 2 &   14 \\
\hline
DoF  &8 & 7 &7\\
\hline
\end{tabular}
\caption{The variables, constraints, and degrees of freedom.}
\end{table}


\end{document}